\documentclass[preprint,preprintnumbers, prd, floatfix,  superscriptaddress,nofootinbib] {revtex4-1}
\usepackage{graphicx}
\usepackage{epsfig}
\usepackage{bm}
\usepackage{amssymb}
\usepackage{float}
\usepackage{amsmath}
\usepackage{dcolumn}
\usepackage{cancel}
\usepackage[colorlinks]{hyperref}
\usepackage[usenames,dvipsnames]{color}
\usepackage{color}
\usepackage{epstopdf}
\usepackage{nccbbb}
\usepackage{ulem}
\hypersetup{
     breaklinks=true,
    pdfstartview={FitH},    colorlinks=true,       % false: boxed links; true: colored links
    linkcolor=blue,          % color of internal links
    citecolor=red,        % color of links to bibliography
    filecolor=magenta,      % color of file links
    urlcolor=blue,           % color of external links
    anchorcolor=green,      % Color for anchor text
    linktocpage=true
}

\def\doi{http://doi.org}

\begin{document}

\title{Constraints on a special running vacuum model}

\author{Chao-Qiang Geng}
\email{geng@phys.nthu.edu.tw}
%\affiliation{Synergetic Innovation Center for Quantum Effects and Applications (SICQEA), 
%Hunan Normal University, Changsha 410081, China}
\affiliation{School of Fundamental Physics and Mathematical Sciences\\Hangzhou Institute for Advanced Study, UCAS, Hangzhou,  310024 China}
\affiliation{International Centre for Theoretical Physics Asia-Pacific, Hangzhou, 310024 China }
\affiliation{Department of Physics, National Tsing Hua University, Hsinchu, Taiwan 300}
\affiliation{National Center for Theoretical Sciences, Hsinchu, Taiwan 300}

\author{Chung-Chi Lee}
\email{lee.chungchi16@gmail.com}
\affiliation{DAMTP, Centre for Mathematical Sciences, University of Cambridge, Wilberforce Road, Cambridge CB3 0WA, UK}

\author{Lu Yin}
\email{yinlu@gapp.nthu.edu.tw}
\affiliation{Department of Physics, National Tsing Hua University, Hsinchu, Taiwan 300}

\begin{abstract}
We study a special running vacuum model (RVM) with $\Lambda = 3 \alpha H^2+3\beta H_0^4 H^{-2}+\Lambda_0$, 
where $\alpha$, $\beta$ and 
$\Lambda_0$ are the model parameters and $H$ is the Hubble one.
This RVM has non-analytic background solutions for the energy densities of matter and radiation,
which can only be evaluated numerically.  From the analysis of the CMB power spectrum and baryon acoustic oscillation along with the prior of $\alpha>0$
to avoid having a negative dark energy density, we find that $\alpha<2.83\times 10^{-4}$ and $\beta=(-0.2^{+3.9}_{-4.5})\times 10^{-4}$ (95$\%$ C.L.).
We show that the RVM fits the cosmological data comparably to the $\Lambda$CDM. In addition, we relate the fluctuation amplitude
$\sigma_8$ to the neutrino mass sum $\Sigma m_\nu$.
\end{abstract}

\maketitle

\section{Introduction}
Since the discovery of the accelerated expanding universe  in 1998~\cite{Riess:1998cb, Perlmutter:1998np}, 
 dark energy has been the most popular scenario to explain this phenomenon~\cite{Copeland:2006wr}. 
 Among the various  theories, the Lambda Cold Dark Matter ($\Lambda$CDM) model is the simplest one
  to reveal  the nature of   our universe, which also  fits well with all  cosmological observational data. Unfortunately, the $\Lambda$CDM model has some theoretical unsatisfactories, such as ``fine-tuning''~\cite{Weinberg:1988cp, WBook} and ``coincidence''~\cite{Ostriker:1995rn, ArkaniHamed:2000tc} problems.

In order to resolve the ``coincidence''  problem, people have proposed various models to improve the cosmological constant of $\Lambda$ 
in the Einstein's equation, such as the running vacuum models (RVMs)~\cite{Ozer:1985ws, Carvalho:1991ut, Lima:1995ea, Basilakos:2018xjp,Perico:2013mna,Sola:2013gha,Shapiro:2001rh,Sola:2013gha,Sola:2014tta,Grande:2011xf,Gomez-Valent:2015pia,Sola:2016jky,Geng:2016fql,Geng:2017apd,Gomez-Valent:2014fda,Sola:2015wwa,Sola:2016ecz}.
In this kind of the models, $\Lambda$, instead of being a constant, is  a function of the Hubble parameter $H$,  and decays to matter and radiation ~\cite{Ozer:1985ws}.
It has been shown that the RVMs are suitable in describing the cosmological evolutions on both background and linear perturbation levels 
in the literature~\cite{Geng:2016dqe,Barrow:2006hia,Shapiro:2009dh,Shapiro:2004ch,Basilakos:2009wi,  Costa:2012xw, Gomez-Valent:2014rxa,  Tamayo:2015qla, Fritzsch:2016ewd,Gomez-Valent:2014fda,Sola:2015wwa,Sola:2016jky,Sola:2016ecz,Geng:2016fql,Geng:2017apd,Zhang:2018wiy,Basilakos:2018xjp,Perico:2013mna}. 
In these studies, the Hubble parameter $H$ has been used to compose many forms of 
 $\Lambda = \sum A_n H^{2n} $ with a non-negative integer n, where $A_n$ is a mass dimension $2(1-n)$ constant.
In this paper, we consider the specific extension of RVM from Ref.~\cite{Basilakos:2018xjp}, in which a negative power term is proposed, 
%{\color{red}
$\Lambda = 3 \alpha H^2+3\beta H_0^4 H^{-2}+\Lambda_0$,  where $\alpha$ and $\beta$ are the model parameters.
% while $\Lambda_0$ plays the role of the cosmological constant.
%In this paper, we concentrate on a particular RVM with $\Lambda = 3 \alpha H^2+3\beta H_0^4 H^{-2}+\Lambda_0$,  where $\alpha$ and $\beta$ are the model parameters, while $\Lambda_0$ is the cosmological constant in the $\Lambda$CDM model.
 Clearly, this RVM goes back to $\Lambda$CDM when $\alpha=\beta=0$. Only in this case, $\Lambda_0$ plays the role of the cosmological constant.
%}
%{\color{pink} The $\Lambda_0$ and $3 \alpha H^2+3\beta H_0^4 H^{-2}$ combined the running vacuum energy $\Lambda$.}
  Naively, it is expected that the values of $\alpha$ and $\beta$ should  be 
  close to zero in order to fit the current cosmological observations.
However, in some of the RVMs, the model parameters have been shown to be non-zero and sizable~\cite{Gomez-Valent:2014fda,Sola:2015wwa,Gomez-Valent:2015pia,Sola:2016jky,Sola:2016ecz,Geng:2016fql,Geng:2017apd}.
It is interesting to explore if our special form of the RVMs also has this peculiar feature.

In this study, we plan to fit this  RVM by using the most recent  observational data.
In particular, we use the {\bf CAMB}~\cite{Lewis:1999bs} and {\bf CosmoMC}~\cite{Lewis:2002ah} packages with the Markov chain Monte Carlo (MCMC) method. Since this model has no analytical solution for the energy density of matter or radiation, we modify the {\bf CAMB} program to get the background evolution. 

This paper is organized as follows.
In Sec.~\ref{sec:model}, we introduce our special RVM. We also derive the evolution equations for matter and radiation in the linear perturbation theory.
In Sec.~\ref{sec:observation}, we present our numerical calculations. In particular, we show
the CMB and matter power spectra and the constraints on the model-parameters from several cosmological observation datasets.
Finally, our conclusions are given in Sec.~\ref{sec:CONCLUSIONS}.

\section{RUNNING VACUUM MODEL}
\label{sec:model}

We start with the Einstein equation, written as
\begin{equation}
\label{eq:action}
R_{\mu\nu}-\frac{1}{2}Rg_{\mu\nu}+\Lambda g_{\mu\nu} = 8\pi G T^{M}_{\mu\nu},
\end{equation}
where $R=g^{\mu\nu}R_{\mu\nu}$ is the Ricci scalar,  $\Lambda$ is the cosmological constant, $G$ is the gravitational constant and $T^{M}_{\mu\nu}$ is the energy-momentum tensor of matter and radiation.
For the homogeneous and isotropic universe, we use the Friedmann-Lemaitre-Robertson-Walker (FLRW) metric, given by
\begin{eqnarray}
ds^{2}= -dt^{2}+ a^{2}(t) \delta_{ij}dx^{i}dx^{j} \,.
\end{eqnarray}
Consequently, the Friedmann equations are found to be
\begin{eqnarray}
\label{eq:friedmann1}
&& H^{2}=\frac{8\pi G}{3}(\rho_m + \rho_r +\rho_{\Lambda}) \,, \\
\label{eq:friedmann2}
&& \dot{H}=- 4\pi G(\rho_m + \rho_r +\rho_{\Lambda} + P_m+ P_r + P_{\Lambda}), \,
\end{eqnarray}
 	where $H=da/(adt)$ is the Hubble parameter and $\rho_{m,r,\Lambda}$ ($P_{m,r,\Lambda}$) represent the energy densities (pressures) of matter, radiation and dark energy, respectively.
In this work,  we consider $\Lambda$ to be the specific function of the Hubble parameter, given by
\begin{equation}
\label{eq:rhode}
\Lambda = 3 \alpha H^2+3\beta H_0^4 H^{-2}+\Lambda_0\,.
\end{equation}
Here,  $\alpha$ and $\beta$ are dimensionless model-parameters. It is clear that 
the $\Lambda$CDM model is recovered by taking $\alpha=0$ and $\beta=0$.
This special model is inspired by the studies of $\Lambda = c_0+c_1H^2+c_2 H^{-n}$ in Refs.~\cite{Perico:2013mna,Basilakos:2018xjp}.
It is convenient to define the  equations of state for matter, radiation and dark energy by
\begin{eqnarray}
\label{eq:eos}
w_{m,r,\Lambda}=\frac{P_{m,r,\Lambda}}{\rho_{m,r,\Lambda}}= 0, \frac{1}{3}, -1 \,,
\end{eqnarray}
respectively.

In the RVM, dark energy decays to radiation and matter in the evolution of the universe,
so that the continuity equations can be written as,
\begin{eqnarray}
\label{eq:continuity}
&& \dot{\rho}_M+3 H(1+w_M)\rho_M = Q \,, \\
&& \dot{\rho}_\Lambda+3 H(1+w_\Lambda)\rho_\Lambda = - Q \,,
\end{eqnarray}
 where
  $\rho_\Lambda = \Lambda/(8\pi G)$, $\rho_M = \rho_m + \rho_r$, $w_M = (P_m + P_r) / \rho_M$ and $Q = Q_m+Q_r$ with $Q_{m(r)}$ the decay rate of dark energy to matter (radiation).
By combining Eqs.~\eqref{eq:rhode} and \eqref{eq:continuity}, the coupling $Q_\mu$ with $\mu$ = $m$ or $r$ is given by
\begin{eqnarray}
\label{eq:coupling}
Q_\mu=-\frac{\dot{\rho}_{\Lambda}(\rho_\mu+P_\mu)}{\rho_{M}+P_M}=3 H \left(\alpha-\beta \frac{H_0^4}{H^4}\right)(1+w_\mu)\rho_\mu \,,
\end{eqnarray}
with $P_M = P_m + P_r$.% where $\mu$ represents matter or radiation.

The energy densities of matter and radiation can be evaluated from 
\begin{eqnarray}
\label{eq:RHOMU}
\frac{\rho_\mu^\prime}{\rho_\mu}= 3(1+w_\mu)\left(\alpha-\beta \frac{H_0^4}{H^4}-1 \right)\,,
\end{eqnarray}
derived from Eq.~\eqref{eq:continuity},
where ``$\prime$" stands for the derivative with respective to $\ln a$ and  $\rho_\mu^\prime = \dot{\rho}_\mu/H$. 
However, there are no analytical solutions for $\rho_{m,r}$ in Eq.~(\ref{eq:RHOMU}).
From the modified ${\bf CAMB}$ program, we can solve  Eq.~\eqref{eq:RHOMU} numerically. 
Note that $\alpha \geq 0$ is chosen to avoid the negative dark energy density in the early universe.

In our calculation, we use the conformal time $\tau$ in order to perform the perturbation theory in the synchronous gauge. From the standard linear perturbation theory~\cite{Ma:1995ey}, we can derive the growth equation of the density perturbation in the RVM.
In the synchronous gauge, the metric  is given by
\begin{eqnarray}
\label{eq:Metric}
ds^{2}=a^{2}(\tau)[-d\tau^{2}+(\delta_{ij}+h_{ij})dx^{i}dx^{j}],
\end{eqnarray}
where  $i,j=1,2,3$ and
\begin{eqnarray}
h_{ij}= \int d^3 k e^{i\vec k\vec x}\left[\hat{k}_i \hat{k}_j h(\vec k,\tau)+6\left(\hat{k}_i\hat{k}_j-\frac{1}{3}\delta_{ij}\right)\eta(\vec k,\tau)\right]\,,
\end{eqnarray}
with  the $k$-space unit vector of $\hat{k}= \vec k/k$   and two scalar perturbations of $h(\vec k,\tau)$ and $\eta(\vec k,\tau)$.
The conservation equation is given by
$\nabla^{\nu}( T^{M}_{\mu\nu}+T^{\Lambda}_{\mu\nu})=0$ with
 $\delta T^{0}_{0}=-\delta \rho_{m}$, 
$\delta T^{0}_{i}=-T^{i}_{0}=(\rho_{M}+P_{M})v^i_M$ and $\delta T^{i}_{j}=\delta P_{M}\delta ^{i}_{j}$.

As shown in Refs.~\cite{DEP2,Grande:2008re}, there are two basic perturbation equations, given by
 \begin{eqnarray}
\label{eq:theta1}
\sum_{i = \Lambda, M} \delta \rho_i+3\delta\frac{\dot a}{a}(\rho_i+P_i)+3\mathcal{H}\left(\delta\rho_i+\delta P_i\right) = 0,
\end{eqnarray}

\begin{eqnarray}
\label{eq:theta2}
\sum_{i = \Lambda, M}\dot\theta_i(\rho_i+P_i)+\theta_i (\dot\rho_i+\dot P_i+5\frac{\dot a}{a}(\rho_i+P_i)) = \frac{k^2}{a}\sum_{i = \Lambda, M}\delta P_i,
\end{eqnarray}
where 
%$\mathcal{H} = da/(a d\tau) = aH$,
 $\delta \rho_i$ represent the density fluctuations,
and $\theta_i$ are the corresponding velocities. As there is no  peculiar velocity for dark energy, we take $\theta_\Lambda=0$.
In addition, we assume that $\delta\rho_M\gg \delta\rho_\Lambda$ and $\delta\dot{\rho}_M \gg \delta\dot{\rho}_\Lambda$  in our model.
As a result, we can ignore the discussion for the dark energy perturbation.
%%%
 For the matter perturbation,
the growth equations are given by
\begin{eqnarray}
\label{eq:pert1}
&& \dot{\delta}_\mu=-(1+w_\mu)\left(\theta_{\mu}+\frac{\dot{h}}{2}\right)-3\frac{\dot a}{a}\left(\frac{\delta P_{\mu}}{\delta \rho_{\mu}}-w_{\mu}\right)\delta_{\mu}-\frac{Q_\mu}{\rho_{\mu}}\delta_{\mu} \,, 
\\
\label{eq:pert2}
&& \dot{\theta}_\mu=-\frac{\dot a}{a}(1-3w_{\mu})\theta_{\mu}-\frac{\dot{w}_{\mu}}{1+w_{\mu}}\theta_{\mu}+\frac{\delta P_{\mu}/\delta \rho_{\mu}}{1+w_\mu}k^{2}\delta_\mu-\frac{Q_\mu}{\rho_{\mu}}\theta_{\mu} \,,
%\\
%\label{eq:pert4}
%&& \dot{\theta}_r=-\mathcal{H}(1-3w_{r})\theta_{r}-\frac{\dot{w}_{r}}{1+w_{r}}\theta_{r}+\frac{\delta P_{r}/\delta \rho_{r}}{1+w_r}\frac{k^{2}}{a^{2}}\delta_r-\frac{Q_r}{\rho_{r}}\theta_{r} \,,
\end{eqnarray}
where $\delta_\mu\equiv \delta \rho_\mu/\rho_\mu$
and  $\mu$ = $m, r$.

\section{Numerical calculations}
\label{sec:observation}

 As mentioned in the previous section, we modify the  ${\bf CAMB}$ program  to solve Eq.~\eqref{eq:RHOMU}. In our calculation, $\rho_m$ and $\rho_r$ are evaluated in terms of  $\log a$ from the current universe to the past. 
By performing the {\bf CosmoMC} program~\cite{Lewis:2002ah},
we fit the RVM from the observational data with the MCMC method.
The dataset includes those of  the
CMB temperature fluctuation from {\it Planck 2015} with TT, TE, EE and low-$l$
polarization~\cite{Adam:2015wua, Aghanim:2015xee, Ade:2015zua}, 
	the  baryon acoustic oscillation (BAO) data from 6dF Galaxy Survey~\cite{Beutler:2011hx,Carter:2018vce},	
	the WiggleZ Dark Energy Survey \cite{Kazin:2014qga} and BOSS~\cite{Anderson:2013zyy,Gil-Marin:2015nqa,Gil-Marin:2018cgo} and the redshift space distortion (RSD) data from SDSS-III BOSS\cite{Gil-Marin:2016wya}. 
	The BAO data points are shown in Table~\ref{tab:1}.
	
In addition, the $\chi^2$ fit is given by
\begin{eqnarray}
\label{eq:chi}
{\chi^2}={\chi^2_{BAO}}+{\chi^2_{CMB}}+{\chi^2_{RSD}}.
\end{eqnarray}
For the BAO, the observation measures the distance ratio of $d_{z}\equiv r_{s}(z_{d})/D_{V}(z)$, where $D_{V}$ is the volume-averaged distance and
$r_{s}(z_d)$ is the comoving sound horizon with $z_{d}$ the redshift at the drag epoch~\cite{Percival2010}.
Here, $D_{V}(z)$ is defined as \cite{Eisenstein2005}
\begin{eqnarray}
	D_{V}(z)\equiv\left[(1+z)^{2}D_{A}^{2}(z)\frac{z}{H(z)}\right]^{1/3},
\end{eqnarray}
	where $D_{A}(z)$ is the proper angular diameter distance, given by
\begin{eqnarray}
	D_{A}(z)=\frac{1}{1+z}\int_{0}^{z}\frac{dz'}{H(z')},
\end{eqnarray}
	while $r_{s}(z)$ is  described  by	
\begin{eqnarray}
	r_{s}(z)=\frac{1}{\sqrt{3}}\int_{0}^{1/(1+z)}\frac{da}{a^{2}H({\scriptstyle z'=\frac{1}{a}-1})\sqrt{1+(3\Omega_{b}^{0}/4\Omega_{\gamma}^{0})a}},\end{eqnarray}
	where $\Omega_{b}^{0}$ and $\Omega_{\gamma}^{0}$ are the present
	values of baryon and photon density parameters, respectively.
The $\chi^2$ value for the BAO data is given by	
	\begin{eqnarray}
	\label{eq:chi}
	{\chi^2_{BAO}}= \sum_{i=1}^n \frac{({D_V/r_s}^{th}(z_i) - {D_V/r_s}^{obs}(z_i))^2}{\sigma^2_i},
	\end{eqnarray}
where $n$ is the number of the BAO data points and $\sigma_i$ correspond to the errors of the data, given by Table~\ref{tab:1}.
% ${D_V/r_s}_{th}(z_i)$ and ${D_V/r_s}_{obs}(z_i)$ 
Here, the subscripts of ``th'' and ``obs'' represent
 the theoretical and observational  values of the volume-averaged distance, respectively.
 
The CMB is sensitive to the distance to the decoupling epoch $z_{*}$. It constrains  the model in the high redshift region of $z\sim 1000$.
The $\chi^{2}$ value of the CMB data can be calculated by
\begin{eqnarray}
\chi_{CMB}^{2}=(x_{i,CMB}^{th}-x_{i,CMB}^{obs})(C_{CMB}^{-1})_{ij}(x_{j,CMB}^{th}-x_{j,CMB}^{obs}),\end{eqnarray}
where $C_{CMB}^{-1}$ is the inverse covariance matrix and
$x_{i,CMB}\equiv\left(l_{A}(z_{*}),R(z_{*}),z_{*}\right)$ with 
the acoustic scale $l_{A}$ and shift parameter $R$, defined by
\begin{eqnarray}
l_{A}(z_{*})\equiv(1+z_{*})\frac{\pi D_{A}(z_{*})}{r_{S}(z_{*})}
\end{eqnarray}
and 
\begin{eqnarray}
	R(z_{*})\equiv\sqrt{\Omega_{m}^{0}}H_{0}(1+z_{*})D_{A}(z_{*}),
\end{eqnarray}
respectively.	

For the RSD measurements, we use
\begin{eqnarray}
{\chi^2_{RSD}}= \sum_{i=1}^n (D^{obs}_{z_i} - D^{th}_{z_i})^\mathrm{T} C^{-1}_{z_i}(D^{obs}_{z_i} - D^{th}_{z_i}),
\end{eqnarray}
\begin{eqnarray}
D_{z}=\left(\begin{array}{c}
f(z)\sigma_8(z) \\
H(z)r_s(z_d) \\
D_A(z)/{r_s({z_d})}
\end{array} 
\right)
\end{eqnarray}
where 
$\sigma_8$ is the amplitude of the over-density at the comoving 8$h^{-1}$ Mpc scale and $f(z)$=$\delta^{\prime}/\delta$ with $\delta$  the evolution of the matter density contrast. 

The data points in RSD are given by 
\begin{eqnarray}
D^{obs}_{0.32}=\left(\begin{array}{c}
0.45960 \\
11.753 \\
6.7443
\end{array} 
\right)~~\text{and}~~
D^{obs}_{0.57}=\left(\begin{array}{c}
0.41750 \\
13.781 \\
9.3276
\end{array} 
\right),
\end{eqnarray}
%\end{eqnarray}
with the covariance matrices being
\begin{eqnarray}
C^{-1}_{0.32}=\left( \begin{array}{ccc}
406.87 & -16.551 & -64.272\\
-16.551 & 6.0291 & -5.6683\\
-64.272 & -5.6683 & 44.018
\end{array} 
\right)~\text{and}~
C^{-1}_{0.57}=\left( \begin{array}{ccc}
1402.2 & -24.384 & -202.70\\
-24.384 & 19.007 & -15.976\\
-202.70 & -15.976 & 95.850
\end{array} 
\right),
\end{eqnarray}
respectively.

\begin{table}[ht]
\begin{center}
	\caption{ BAO data points.  }
	\begin{tabular}{|c|c|c|c||c|c|c|c|}
		\hline
		\ & $z$ & BAO($D_V/r_s$) & Ref. & \ & $z$ & BAO($D_V/r_s$) & Ref. \\
		\hline
		~$1$~ & $0.097$ & $2.52 \pm 0.12$ &\cite{Carter:2018vce} &
		~$6$~ & $0.44$ & $11.57 \pm 0.56$ &\cite{Kazin:2014qga}\\
		\hline
		~$2$~ & $0.106$ & $2.976 \pm 0.176$ & \cite{Beutler:2011hx} &
		~$7$~ & $0.56$ & $13.70 \pm 0.12$ & \cite{Gil-Marin:2015nqa}\\
		\hline
		~$3$~ & $0.15$ & $4.47 \pm 0.17$ & \cite{Anderson:2013zyy} &
		~$8$~ & $0.60$ & $14.98 \pm 0.68$ & \cite{Kazin:2014qga} \\
		\hline
		~$4$~ & $0.122$ & $3.65 \pm 0.12$ &\cite{Carter:2018vce} &
		~$9$~ & $0.73$ & $16.97 \pm 0.58$ &\cite{Kazin:2014qga}\\
		\hline
		~$5$~ & $0.32$ & $8.62 \pm 0.15$ & \cite{Gil-Marin:2015nqa}&
		\ & \ & \ & \
		\\
		\hline
		\hline
		\ & $z$ & BAO($D_A/r_s$) & Ref. & \ &  & & \\
		\hline
%		~$1$~ & $0.32$ & $6.6 \pm 0.13$ &\cite{Gil-Marin:2016wya} &
%		~$2$~ & $0.57$ & $9.39 \pm 0.1$ & \cite{Gil-Marin:2016wya} \\
%		\hline
		~$3$~ & $1.52$ & $12.48 \pm 0.71$ & \cite{Gil-Marin:2018cgo} &
		\ & \ & \ & \
		\\
		\hline
		%						\hline
		% \ & $z$ & RSD($D_A/r_s$) & Ref. & \ & $z$ & RSD($D_A/r_s$) & Ref.  \\
	\end{tabular}
	%\vskip 0.2in
	\label{tab:1}
\end{center}
\end{table}
%In addition, the $\chi^2$ fit is given by
The priors of the various 
cosmological parameters are listed in Table~\ref{tab:3}.
Here,  we have set $\alpha$ to be a positive number to avoid having
a negative dark energy density. 

\begin{table}[ht]
\begin{center}
\caption{ Priors for cosmological parameters with $\Lambda= 3\alpha H^2 + 3\beta H_0^4/H^2 + \Lambda_0$.  }
\begin{tabular}{|c||c|} \hline
Parameter & Prior
\\ \hline
Model parameter $\alpha$& $0 \leq 10^{4}\alpha \leq  10$
\\ \hline
Model parameter $\beta$& $-10 \leq 10^{4}\beta \leq 10$
\\ \hline
Baryon density parameter& $0.5 \leq 100\Omega_bh^2 \leq 10$
\\ \hline
CDM density parameter & $0.1 \leq 100\Omega_ch^2 \leq 99$
\\ \hline
Optical depth & $0.01 \leq \tau \leq 0.8$
\\ \hline
Neutrino mass sum& $0 \leq \Sigma m_{\nu} \leq 2$~eV
\\ \hline
$\frac{\mathrm{Sound \ horizon}}{\mathrm{Angular \ diameter \ distance}}$  & $0.5 \leq 100 \theta_{MC} \leq 10$
\\ \hline
Scalar power spectrum amplitude & $2 \leq \ln \left( 10^{10} A_s \right) \leq 4$
\\ \hline
Spectral index & $0.8 \leq n_s \leq 1.2$
\\ \hline
\end{tabular}
%\vskip 0.2in
\label{tab:3}
\end{center}
\end{table}

\begin{figure}
\centering
\includegraphics[width=0.45 \linewidth, angle=270]{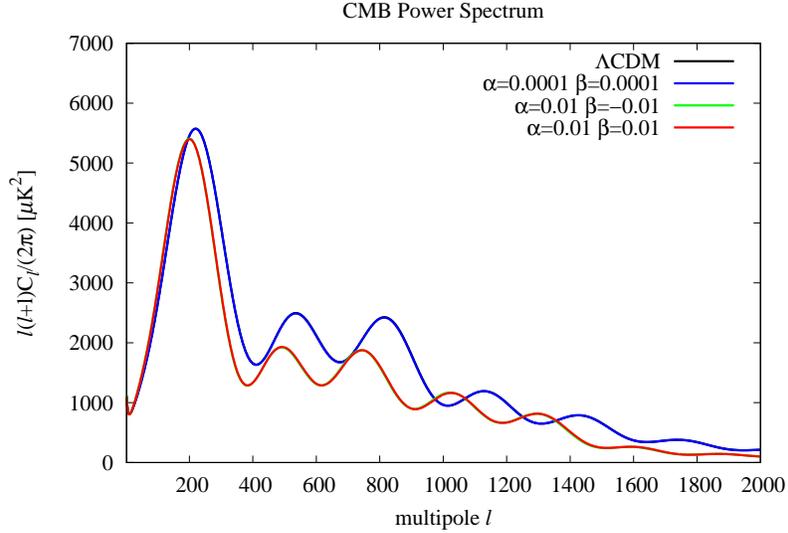}
\caption{CMB power spectra for the $\Lambda$CDM and RVM with different sets of $\alpha$ and $\beta$,  where $\alpha$=0.0001 and $\beta$=0.0001 in  the RVM are illustrated by the blue line, which almost coincides with the black one from  $\Lambda$CDM,
while  the green and red lines with $\alpha$=0.01 are also almost in the same line. }
\label{fg:1}
\end{figure}

\begin{figure}
	\centering
	\includegraphics[width=0.45 \linewidth, angle=270]{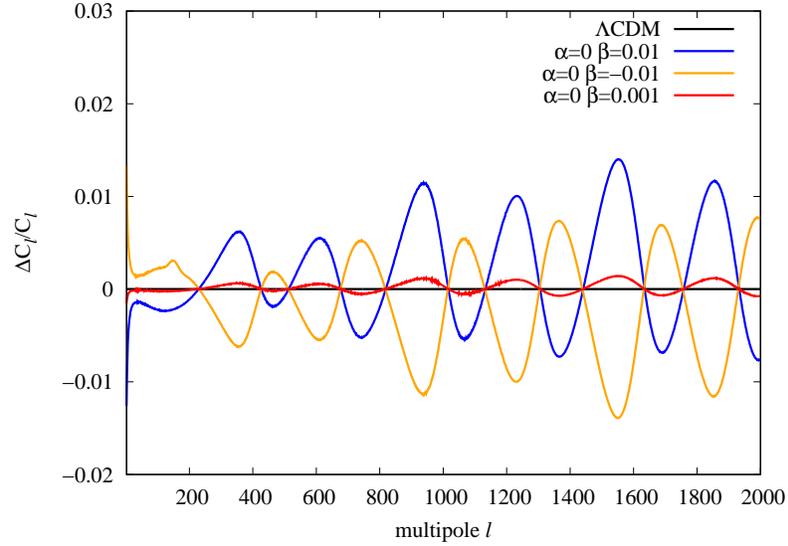}	
	\caption{
	Ratio of  $\Delta C_{\ell}/C_{\ell}$, where $\Delta C_{\ell}$ is the change 
 between the RVM and $\Lambda$CDM for the TT mode of the CMB  power spectra, while $C_{\ell}$ corresponds to the one in $\Lambda$CDM.
} %\caption{CMB power spectrum for $\Lambda CDM$ and RVM in
	% different $\alpha$ and $\beta$.}
	\label{fg:4}
\end{figure}

\begin{figure}
\centering
\includegraphics[width=0.45 \linewidth, angle=270]{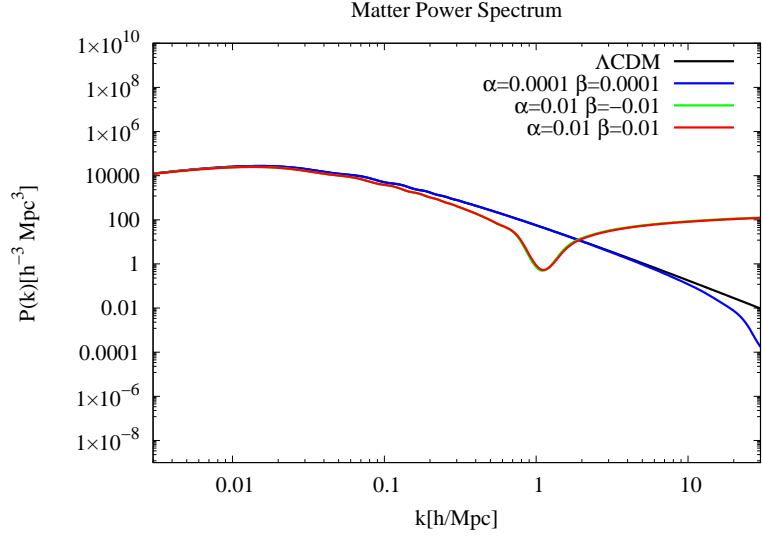}
\caption{Matter power spectra for the  $\Lambda$CDM and RVM, where the legend is the same as Fig.~\ref{fg:1}.}
\label{fg:2}
\end{figure}

In Fig.~\ref{fg:1}, we show the CMB  power spectra in the  $\Lambda$CDM and RVM 
with several different sets of $\alpha$ and $\beta$. 
In the figure, we see that both model parameters $\alpha$ and $\beta$ in Eq.~\eqref{eq:rhode} are expected to be smaller than 0.0001 
as illustrated by the blue line, which almost coincides with the $\Lambda$CDM one (black).
In the green and red lines with $\alpha$=0.01, the first acoustic peaks are reduced, which could result from
 too much contribution from dark energy to the total energy density to suppress the baryon part~\cite{Stadler:2018dsa}.
  In this case, the overall shift of the Doppler peaks towards lower multipoles as a consequence  of the increased sound speed of the plasma~\cite{Stadler:2018dsa}. Compared with $\alpha$, the effect of $\beta$ in the CMB is not obvious, because $H(z)$ is very large due to the $H^2$ term in the early universe.
 In Fig.~\ref{fg:4}, we give the ratio of $\Delta C_{\ell}/C_{\ell}$, where $\Delta C_{\ell}$ is the change 
 between the RVM and $\Lambda$CDM for the TT mode of the CMB  power spectra, while $C_{\ell}$ corresponds to the one in $\Lambda$CDM.
This figure illustrates the effects from the model-parameter of $\beta$ from -0.01 to 0.01. 
It is clear that  the changes in the CMB power spectra due to $\beta$ are small, so that
the results in the RVM will be only slightly different  from those in $\Lambda$CDM  in the early universe.
We present the  matter power spectra of the RVM in Fig.~\ref{fg:2}, which behavior
similar to those in Figs.~\ref{fg:1} and \ref{fg:4}.
In addition, 
%we demonstrate  that $\rho_{\Lambda}$ is dominated by the term of $\alpha H^2$ in the early universe. 
%In Fig.~\ref{fg:2}, 
we demonstrate  that  the matter power spectra for the RVM and $\Lambda$CDM do not have similar evolution paths in the early universe until k$\approx$5.

\begin{figure}
\centering
\includegraphics[width=0.96 \linewidth]{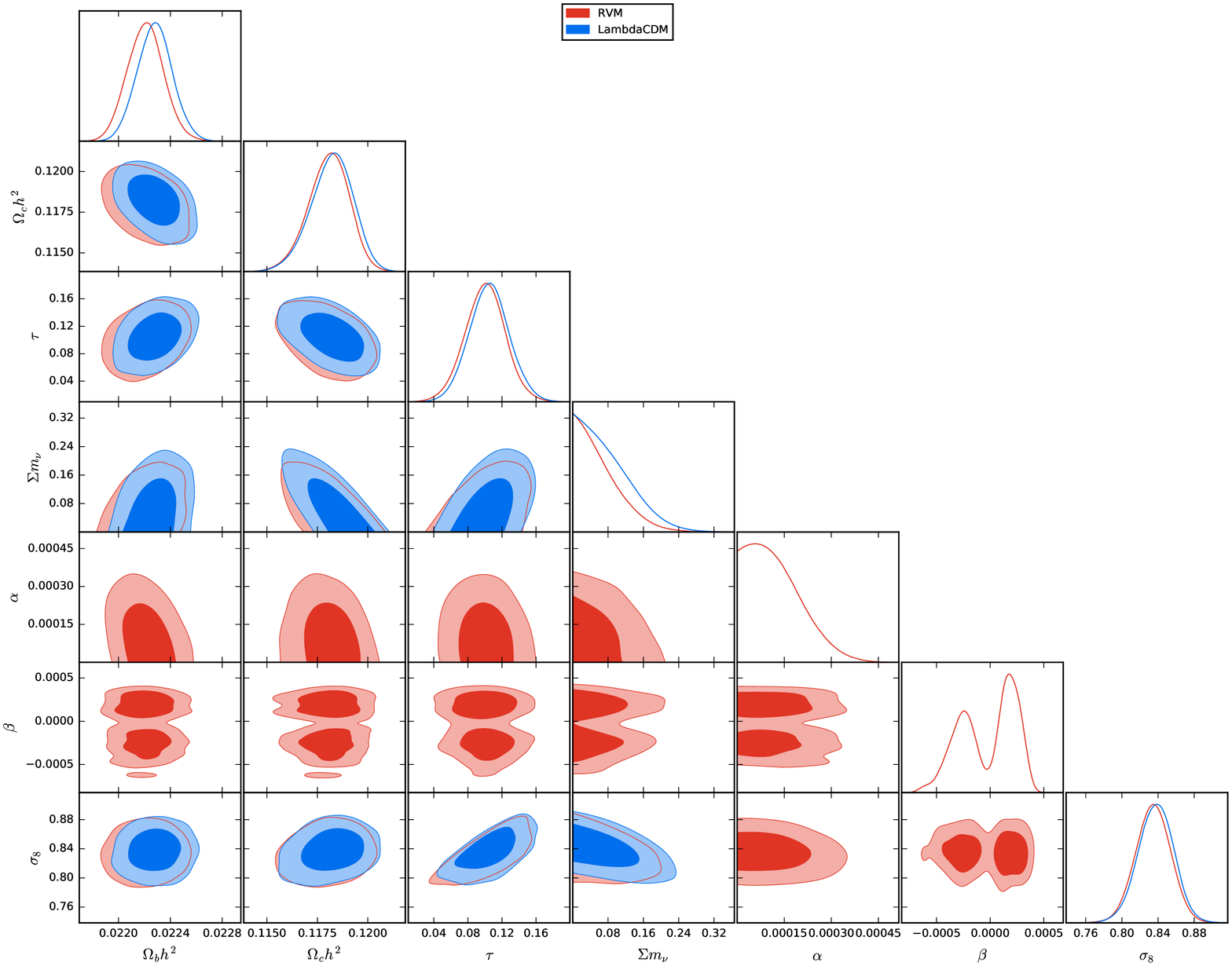}
\caption{One and two-dimensional distributions of $\Omega_b h^2$, $\Omega_c h^2$, $\tau$, $\sum m_\nu$, $\alpha$, $\beta$, $\sigma_8$, where the contour lines represent 68$\%$~ and 95$\%$~ C.L., respectively.}
\label{fg:3}
\end{figure}

\begin{table}[ht]
	\begin{center}
		\caption{Fitting results for the RVM and $\Lambda$CDM, where the limits are given at 68$\%$ and 95$\%$ C.L., respectively }
		\begin{tabular} {|c|c|c|c|c|}
			\hline
			Parameter & RVM (68\% C.L.)& RVM (95\% C.L.) & $\Lambda$CDM (68\% C.L.)& $\Lambda$CDM (95\% C.L.)\\
			\hline
			{\boldmath$\Omega_b h^2   $} &$0.02221\pm 0.00014       $& $0.02221^{+0.00028}_{-0.00027}$& $0.02228\pm 0.00014        $& $0.02228^{+0.00027}_{-0.00026}$\\
			
			{\boldmath$\Omega_c h^2   $} & $0.1181^{+0.00110}_{-0.00090}          $& $0.1181^{+0.00190}_{-0.00210}$& $0.1182^{+0.00110}_{-0.00094}          $& $0.1182^{+0.00190}_{-0.00220}$\\
			
		%{\color{red}
		 {\boldmath$\Omega_{\Lambda} $}& $0.6830^{+0.0085}_{-0.0072} $
		& $0.6830^{+0.0149}_{-0.0130}$  & $0.6813^{+0.0131}_{-0.0022} $ & $0.6813^{+0.0193}_{-0.0116}$\\
			
			{\boldmath$100\theta_{MC} $} &$1.04122\pm 0.00032       $&$1.04122\pm{0.00062}$& $1.04108\pm 0.00030      $& $1.04108^{+0.00060}_{-0.00059}$\\
			
			{\boldmath$\tau           $} & $0.100\pm 0.024           $& $0.100\pm{0.047}   $&  $0.105\pm 0.023           $& $0.105^{+0.047}_{-0.045}   $\\
			
			{\boldmath$\Sigma m_\nu   $} & $< 0.0774                   $& $< 0.160                   $& $< 0.0993                   $& $< 0.186                    $\\
			
			{\boldmath$10^4\alpha         $} & $< 1.57               $& $< 2.83                $&-&-\\
			
			{\boldmath$10^4\beta          $} & $-0.2\pm 2.6$& $-0.2^{+3.9}_{-4.5}$&-&-\\
			
			{\boldmath${\rm{ln}}(10^{10} A_s)$} & $3.134\pm 0.046           $& $3.134^{+0.090}_{-0.092}   $& $3.142\pm 0.046           $& $3.142^{+0.091}_{-0.088}   $\\
			
			$H_0                       $ & $66.66^{+0.45}_{-0.40}            $& $66.66^{+0.79}_{-0.85}         $&$66.92\pm 0.40         $& $66.92^{+0.76}_{-0.80}        $\\
			
			$\sigma_8                  $ & $0.835\pm 0.019     $&$0.835\pm{0.038}   $&$0.838\pm 0.020    $& $0.838^{+0.038}_{-0.040}   $\\
			\hline
			$\chi^2_{best-fit} $& \multicolumn{2}{c|}{2543.259}& \multicolumn{2}{|c|}{2546.662}\\
			\hline
		\end{tabular}
		\label{tab:2}
	\end{center}
\end{table}

\begin{figure}
\centering
\includegraphics[width=0.80 \linewidth]{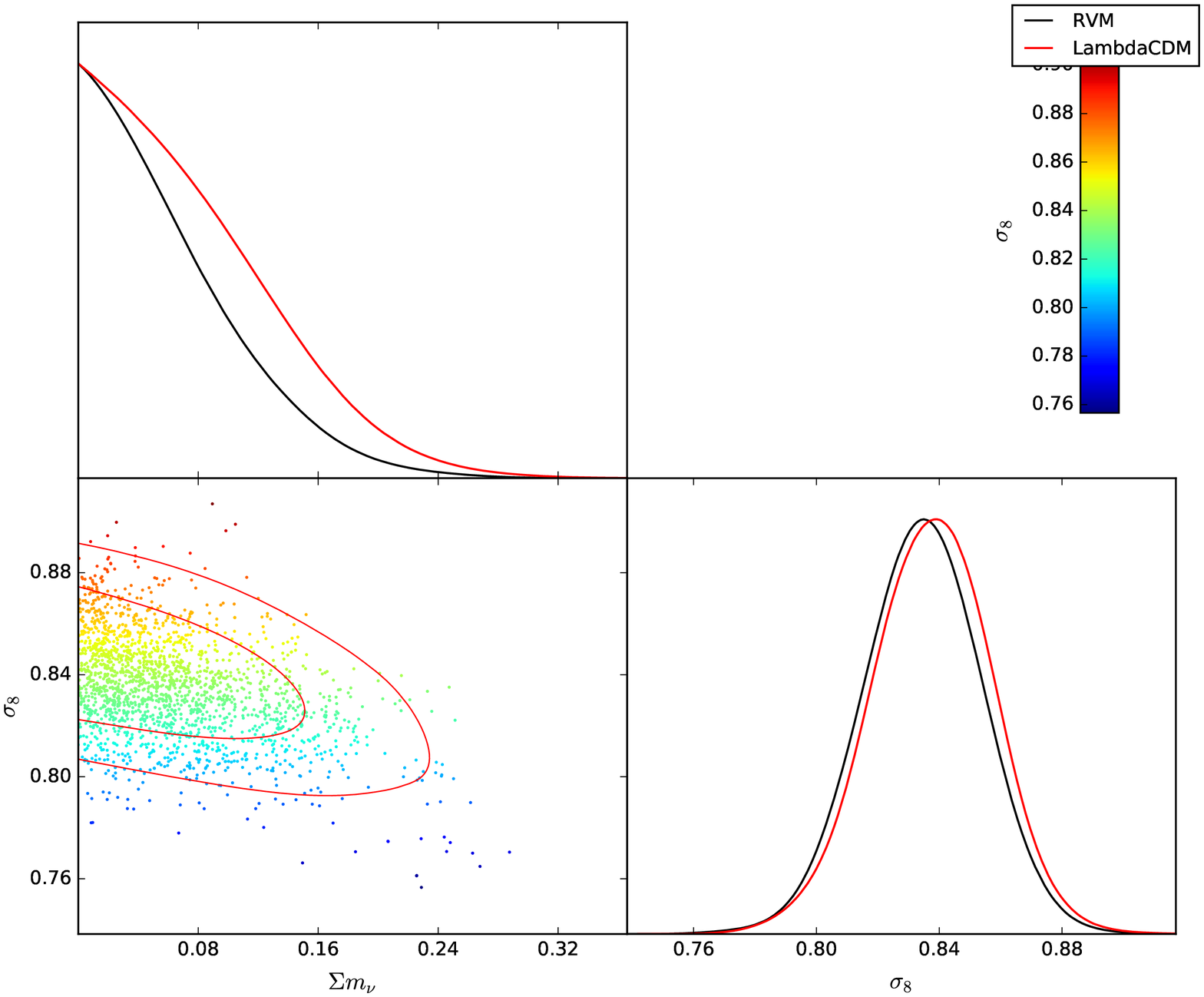}
\caption{One and two-dimensional distributions of $\Sigma m_\nu   $ and $\sigma_8$, where the contour lines represent 68$\%$~ and 95$\%$~ C.L., respectively.}
\label{fg:5}
\end{figure}

In Fig.~\ref{fg:3} and Table~\ref{tab:2}, we present our results of the global fits from  several  datasets, where the values in the brackets
correspond to  the best-fit values in the $\Lambda$CDM model.
In particular, we find that ($\alpha$, $\beta$) = $(<2.83, -0.2^{+3.9}_{-4.5})\times 10^{-4}$ and $(<1.57, -0.2\pm 2.6)\times 10^{-4}$ 
with 95$\%$ and 68$\%$~C.L., respectively.
%{\color{red}
Here, $\Omega_{\Lambda}=( \rho_{\Lambda}/\rho_C)|_{z=0} = \alpha+\beta+\Lambda_0/(3H_0^2)$ 
is the fractional dark energy density
with $\rho_{\Lambda}\equiv \Lambda/(8\pi G)$ and $\rho_{C}= 3H^2/(8\pi G)$.
%Taking $\alpha$=0 and $\beta$=$-2\times10^{-5}$, the central value in Table~\ref{tab:2}, one can have $\Lambda_0/(3H^2_0)=0.6830$.
%}
It is interesting to note that the value of $\sigma_8$=$0.835\pm {0.038}  $ (95$\%$~C.L.) in the RVM
is smaller than that of $0.838^{+0.038}_{-0.040}$ (95$\%$~C.L.) in $\Lambda$CDM. 
As shown in  Table~\ref{tab:2}, the best fitted $\chi^2$ value in the RVM is 2543.259, which is  smaller than 2546.662 in the $\Lambda$CDM model.
  Although the cosmological observables for the best $\chi^2$ fit in the RVM do not significantly deviate from those in $\Lambda$CDM, they look better in all datasets.
   It implies that the RVM is favored by the cosmological observations.
 However, we remark that our results can only be viewed as comparable to those in $\Lambda$CDM
 due to the extra parameters in the RVM.
  In addition,
it should be noted that $\alpha\sim \mathcal{O}(10^{-4})$ in our RVM is about one to two orders of magnitude lower than those of
$\alpha \sim \mathcal{O}(10^{-3})- \mathcal{O}(10^{-2})$ of the corresponding $H^2$ term in the other RVMs in the literature~\cite{ Sola:2016jky, Gomez-Valent:2014fda, Sola:2015wwa, Gomez-Valent:2015pia, Sola:2016ecz}.
%{\color{red}
However, this difference may be due to the fact that the models are actually different as there is no term proportional to 
negative powers of $H$ in the cases of the other authors.
%}

Another interesting result is about the correlation between the fluctuation amplitude $\sigma_8$ and the neutrino mass sum  $\Sigma m_\nu$
shown in Fig.~\ref{fg:5}.
Many local observations~\cite{Battye:2014qga,McCarthy:2017csu,Abbott:2017wau} have claimed that the value of $\sigma_8$ should be smaller than the one given by the  Planck measurement~\cite{Ade:2015xua,Aghanim:2018eyx}. 
We remark that our fitted value of $\sigma_8$ is much higher than those in Refs.~\cite{Ade:2015xua,Aghanim:2018eyx} due to the different data set.
It is known that the
 cosmic shear data are important as well and  several surveys usually provide values of $\sigma_8$ much smaller than those from the Planck data~\cite{Joudaki:2019pmv}. Nevertheless, we would examine the tendency of  $\sigma_8$ in our model.
In order to reduce $\sigma_8$, we should have a smaller matter amplitude, which is consistent with our fitting result in Fig.~\ref{fg:3}, in which the RVM has lower values of $\Omega_b$ and $\sigma_8$ than those in the $\Lambda$CDM model. In Fig.~\ref{fg:5} we focus on the relationship between  $\sigma_8$ and $\Sigma m_\nu$, where the red to blue points represent different values of $\sigma_8$ form 0.9 to 0.76. It is clear that a smaller value of $\sigma_8$ allows of a larger $\Sigma m_\nu$. 

\begin{figure}[ht]
\centering
\includegraphics[width=0.6  \linewidth, angle=270]{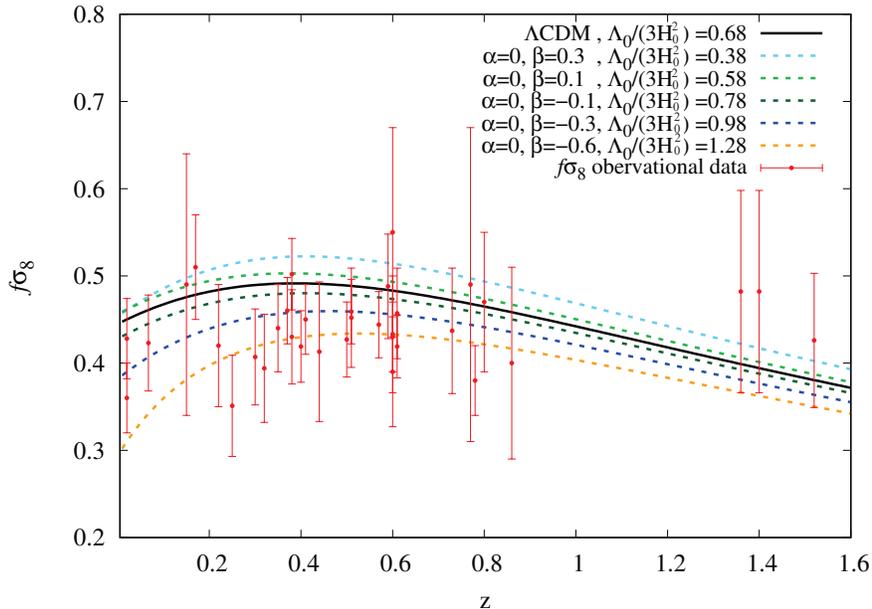}
\caption{$f\sigma_8$ as a function of $z$ in our model and $\Lambda$CDM, 
	%where the dark energy density in each case is fixed to be {\color{red} $\Lambda=6.345\times 10^{-27} kgm^{-3}$ in present time and corresponds to the $\Lambda$ in Eqs.~\eqref{eq:rhode}. 
	% {\color{red}
	 where $\Omega_{\Lambda}$ is fixed to be 0.68.}
\label{fg:6}
\end{figure}

%{\color{red}
Finally, it is interesting to discuss a specific case with $\Lambda_0 = 0$, which might give us a late-time accelerating epoch at the present time and leads our universe to end up with the de-Sitter space in the far future. Such a future dark energy dominated universe can be discussed by substituting Eq.~\eqref{eq:rhode} into Eq.~\eqref{eq:friedmann1} with $\rho_m = \rho_r =0$, that is
\begin{eqnarray}
H^2 = \alpha H^2 + \beta H_0^4 H^{-2} \,,
\end{eqnarray}
which leads to $H^2 = H_0^2 \sqrt{\beta / (1-\alpha)}$, pointing out that the existence of the de-Sitter space appears only if $\beta/(1-\alpha) > 0$. As we have discussed~\cite{Geng:2017apd,Geng:2016fql}, $\alpha \sim 1 $ gives us a large abundance of the dark energy density in the early universe, so that the observations require the value of $\alpha$ to be small. On the other hand, the negative value of $\alpha$ induces $\rho_{\Lambda}<0$ at a high $z$, which should be avoided. %{\color{red} We note that $\rho_{de}=\Lambda/(8\pi G)$}. 
Thus, the allowed window for $\alpha$ is tiny with $ 1 \gg \alpha \geq 0$. As a result, the de-Sitter space can exist if we have a suitable positive value of $\beta \sim \mathcal{O}(1)$. This special case not only keeps the late-time accelerating universe but also further reduces the model parameters by one, i.e., $\Lambda_0 =0$. We show the evolution of $f \sigma_8$ in Fig.~\ref{fg:6} with $\alpha=0$ and different values of $\Lambda_0$ and $\beta$. 
%{\color{red}$\Lambda_0(\beta=0)=6.345\times 10^{-27} kgm^{-3}$ comes from the result of Planck. In order to keep $\Lambda$ in Eqs.~\eqref{eq:rhode} to be a constant number in present time, we set $\Lambda_0(\beta=0.3)=3.716\times 10^{-27} kgm^{-3}$, $\Lambda_0(\beta=0.1)=5.468\times 10^{-27} kgm^{-3}$, $\Lambda_0(\beta=-0.1)=7.221\times 10^{-27} kgm^{-3}$, $\Lambda_0(\beta=-0.3)=8.974\times 10^{-27} kgm^{-3}$ and $\Lambda_0(\beta=-0.6)=1.1603\times 10^{-26} kgm^{-3}$ in Fig.~\ref{fg:6}, respectively.} 
%{\color{red}
Here, to illustrate the behavior of $f\sigma_8$, we have fixed $\Omega_{\Lambda}$  to be the best fitted value
of $0.68$ shown in Table~\ref{tab:2}.
%}
As we can see, the smaller value for $\Lambda_0$ is, the more significant deviation from that of the $\Lambda$CDM prediction behaves, indicating that the vanishment of the $\Lambda_0$ term in the specific RVM does not work well at the linear perturbation level. This result is clearly due to the strong interaction between matter and dark energy in the late time of the universe. A bunch of relativistic and non-relativistic matter decay into dark energy, which further enhance the matter density perturbation $\delta_M$ and change $f \sigma_8$ in our universe. Therefore, this specific case with $\Lambda_0=0$ is available in the background evolution history but, of course, unacceptable at linear perturbation observations.
%}

%{\color{red}
%Finally, we discuss other forms of $\Lambda$ in Eq.~(\ref{eq:rhode}).
 %If $\Lambda_0=0$ in Eq.~(\ref{eq:rhode}), we will not be able to get the $\Lambda$CDM behavior in the large curvature regime,
%which is clearly unacceptable. Specifically,
%the term  of $\Lambda = 3\beta H_0^4 H^{-2}$ alone in Eq.~(\ref{eq:rhode}) can be ignored in the large curvature regime, but it contributes 
%too much dark energy at the present, so that it must be ruled out. 
%On the other hand,  with the term of $\Lambda = 3 \alpha H^2$ alone in Eq.~(\ref{eq:rhode}) to
%fit with the local measurement, it  leads to a too large $\rho_{de}$ in the early universe.
%Clearly, the models with just one parameter do not work in our universe.
%For $\alpha$=0  and $\Lambda_0\neq 0$, the term of $3\beta H_0^4 H^{-2}$ in $\Lambda$
%can be ignored in the early universe, so that the model behaves  just like $\Lambda$CDM. 
%Since this term has an effect in the local universe,  we compare our model with the local observation of $f \sigma_8$ as shown in Fig.~\ref{fg:6}. 
%By taking the values of $\rho_{de}^0$
%%=5.96\times 10^{-27} kg/m^3$ 
%and $H_0$
%%=67.66\pm 0.42 km/s/Mpc=2.19277\pm 0.0136 \times 10^{-18} s^{-1}$ (
%from the data of Planck 2015,  $\rho_{de} = 3\beta H_0^4 H^{-2}+\rho_0^{de}$ will always be positive when $\beta > -4.132\times 10^8$. 
%When we consider $\beta=0$ and $\Lambda_0\neq 0$, the result can be seen in our previous work at Ref.~\cite{Geng:2017apd}.
%%}

\section{Conclusions}
\label{sec:CONCLUSIONS}

We have studied the  RVM with $\Lambda = 3 \alpha H^2+3\beta H_0^4 H^{-2}+\Lambda_0$ . By modifying 
the program in ${\bf CAMB}$, we have solved the  equations for the energy densities of matter and radiation
and obtain the numerical solutions. In the CMB and matter power spectra,  we have used several different sets of 
 $\alpha$ and $\beta$ to show the cosmological evolutions of the model in the early universe. 
 With the data of BAO, RSD and CMB, we have found that $\alpha$ and $\beta$
 are $(<2.83, -0.2^{+3.9}_{-4.5})\times 10^{-4}$ (95$\%$~C.L.) and $(<1.57, -0.2\pm 2.6)\times 10^{-4}$ (68$\%$~C.L.), 
 respectively. The best fitted $\chi^2$ value is 2543.259 in the RVM, which is in the same order but a little smaller than 2546.662 in the $\Lambda$CDM model.
 In addition, the fitting result of $\sigma_8$ has been found to be also smaller than that in $\Lambda$CDM. 
 The results in the  RVM are comparable to those in  $\Lambda$CDM to explain the observational data, especially consistent with the local data in the $\sigma_8$ problem. 
 %Furthermore, we have related $\sigma_8$ and $\Sigma m_\nu$.
 %In particular, we have illustrated that a smaller value of $\sigma_8$ allows of a larger $\Sigma m_\nu$.

\section*{Acknowledgments}
The work was supported in part by National Center for Theoretical Sciences, MoST (MoST-107-2119-M-007-013-MY3),  and the Newton International Fellowship (NF160058) from the Royal Society (UK).

\end{document}